\documentstyle[12pt]{article}

\bigskip
\title{ On the Three-dimensional Lattice Model }
\author{Zhan-Ning Hu \thanks{\bf email address: huzn@itp.ac.cn}
\\CCAST(World Lab.), P.O.Box 8730, Beijing 100080 \\
\\and\\
\\ Institute of Theoretical Physics, Academia Sinica \\
  P. O. Box 2735, Beijing 100080, China \thanks{\bf mail address}}
\date{Jan. 14, 1995}
\begin{document}
\maketitle
\bigskip
\begin{abstract}
Using the restricted star-triangle relation, it is shown that the $N$-state
spin
integrable model on a three-dimensional lattice with spins interacting round
each
elementary cube of the lattice proposed by Mangazeev, Sergeev and Stroganov is
a
particular case of the Bazhanov-Baxter model.

\bigskip

\bigskip

\bigskip


\smallskip

\bigskip

\end{abstract}
\newpage

\section{Introduction}
As a generalization of Zamolodchikov model with the $Z_2$ symmetry, the
interaction-round-cube (IRC) model with the $Z_N$ symmetry was first proposed
by
Bazhanov and Baxter \cite{BB1}. In that model the tetrahedron equation
 \cite{Zam,Jae} plays an important role, by which the commutativity of
 layer-to-layer transfer matrices is ensured in the three-dimensional lattice
model. Later the three-dimensional star-star relation and the tetrahedron
 equation were discussed in Refs. \cite{BB2,Kae1,Kae2,Hu1,Hu2}.

Korepanov investigated  the solution of the  vertex tetrahedron equation in
Ref. \cite{Kor}. The elliptic solutions for modified tetrahedron equations
related to the 3D integrable models  were studied by Boos $et~al$ \cite{Boo}.
Bellon $et~al$ discussed the cubic lattice model by imposing some restricted
 conditions on the entries of the $R$-matrix \cite{BelVia}.

Very recently in Ref. \cite{Man}, another IRC model has proposed in which the
weight function is a vertex solution of the tetrahedron equation with the
arbitrary number $N$ of the spin variables as a generalization of Hietarinta's
result. This solution has the multiplicative form and can be written as
\begin{equation}  \label{1}
R^{j_1j_2j_3}_{i_1i_2i_3}=\omega^{j_1(i_3-j_3)}
{\displaystyle w(p_2,p_{12},p_1|i_3-j_3)w(p_4,p_{34},p_3|i_1-j_1)\over
 \displaystyle w(p_6,p_{56},p_5|j_2-i_2)}
\end{equation}
where the spin variables $i_1$, $i_2$, $i_3$, $j_1$, $j_2$, $j_3$ satisfy the
conditions $j_2=i_1+i_3,~i_2=j_1+j_3$ and take their values in $Z_N$. And the
functions $w(p_2,p_{12},p_1|i_3-j_3), \cdots$ have the following form:
\begin{equation}                                                 \label{2}
w(x,y,z|l)=\prod_{j=1}^{l}{y\over z-x\omega^j},~~x^N+y^N=z^N,
\end{equation}
with the notations
\begin{equation}
\omega=\exp(2\pi i/N),\qquad \omega^{1/2}=\exp(\pi i/N).
\end{equation}
The weight function $R$ satisfies the vertex type tetrahedron equation
\cite{Hir}
\begin{equation}
{\displaystyle\sum_{k_1,k_2,k_3,\atop k_4,k_5,k_6}}
R^{k_1,k_2,k_3}_{i_1,i_2,i_3}R'^{j_1k_4k_5}_{\phantom{,}k_1i_4i_5}
R''^{j_2j_4k_6}_{\phantom{,,}k_2k_4i_6}
R'''^{j_3j_5j_6}_{\phantom{,,,}k_3k_5k_6}
={\displaystyle\sum_{k_1,k_2,k_3,\atop k_4,k_5,k_6}}
R'''^{k_3,k_5,k_6}_{\phantom{,,,}i_3,i_5,i_6}
R''^{k_2k_4j_6}_{\phantom{,,}i_2i_4k_6}
R'^{k_1j_4j_5}_{\phantom{,}i_1k_4k_5}
R^{j_1j_2j_3}_{k_1k_2k_3}.
\end{equation}
It ensures the commutativity of layer-to-layer transfer matrices constructed
from
 weight functions $R$ and $R'$ \cite{Jae,Zam}. In view of this, a natural
question may be raised. What is the difference between the above model and the
Bazhanov-Baxter model? Since the latter has the more parameters in the weight
 function, we may ask if the former one is a special case of the latter model.
So far this question has only been answered in an affirmative way for the case
of $N=2$ (See Ref. \cite{Man}). The aim of  this letter is to give an
affirmative
answer for the generic case of arbitrary number of spin variables.

The outline of this letter is as follows. The Bazhanov-Baxter model is
described
 in $\S$2. In section 3, the weight function denoted by Eq. (1) is derived from
 the Bazhanov-Baxter model. This demonstrates that the former model is the
 particular case of that of latter. Finally, some remarks are given.

\section{The Bazhanov-Baxter Model}

Following Ref. \cite{BB1}, consider a simple cubic lattice $\cal L$. At each
site
 of it has a spin $\sigma \in Z_N$ so that all possible interaction is allowed
 within each elementary cube (See Fig.1). The partition function of the
 Bazhanov-Baxter model can be given from

\begin{picture}(300.0,150.0)(-10,0)
\thicklines
\put(180,5){\circle*{15}}
\put(180,5){\line(1,0){80.0}}
\put(260,5){\circle*{15}}
\put(180,5){\line(0,1){80.0}}
\put(180,85){\circle*{15}}
\put(180,85){\line(1,0){80.0}}
\put(260,85){\circle*{15}}
\put(260,5){\line(0,1){80.0}}
\put(180,5){\line(-2,1){54}}
\put(128.5,30.8){\circle*{15}}
\put(128.5,30.8){\line(0,1){80.0}}
\put(128.5,110.8){\circle*{15}}
\put(180,85){\line(-2,1){54}}
\put(128.5,110.8){\line(1,0){80.0}}
\put(208.5,110.8){\circle*{15}}
\put(260,85){\line(-2,1){54}}
\put(208.5,30.8){\circle*{15}}
\multiput(260,5)(-17.9,8.9){3}{\line(-2,1){10.0}}
\multiput(128.5,30.8)(20,0){4}{\line(1,0){10.0}}
\multiput(208.5,30.8)(0,20){4}{\line(0,1){10.0}}
\put(162, 0){e}
\put(162,80){a}
\put(275,0){d}
\put(275,80){f}
\put(112.5,35.8){c}
\put(111.5,115.8){g}
\put(223.5,115.8){b}
\put(223.5,35.8){h}
\end{picture}

\bigskip
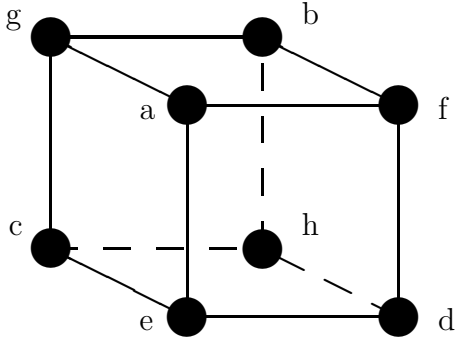
\centerline{Fig.1. {\it Arrangement of the spins $a,b,c,d,e,f,g,,h$ on the
elementary cube}}
\bigskip
\noindent
the following weight function $V$ \cite{BB1}:
$$
V(a|efg|bcd|h)                                    ~~~~~~~~~~~~~~~~~~~~~~~~~~~~~
{}~~~~~~~~~~~~~~~~~~~~~~~~~~~~~~~~~~~~~~
$$$$
=w_{p'p}(e-c-d+h)w^{-1}_{p'p}(a-g-f+b)s(c-h,d-h) ~~~
$$$$
{}~~~~ \times s(g,a-g-f+b)\sum^{N-1}_{\sigma=0}w^{-1}_{p'q}(e-c-\sigma)w_{pq}
(d-h-\sigma)
$$
\begin{equation}     \label{2.1}
\times w_{q'p}(\sigma-f+b) \widetilde{w}_{p'q'}(a-g-\sigma)s(\sigma,a-c-f+h),
\end{equation}
where $w^{-1}_{pq}(l)$ denotes $1/w_{pq}(l)$, and
$$
w_{pq}(k)=w(p/q,k),~~\widetilde{w}_{pq}(k)=\Phi(k)w(p/q,k),
$$
\begin{equation}                                           \label{5}
\Phi(k)=(\omega^{1/2})^{k(N+k)},~~~ s(k,l)=\omega^{kl},
\end{equation}
$$
{w(p/q,l)\over w(p/q,0)}=[\Delta(p/q)]^l\prod^l_{k=1}(1-\omega^kp/q)^{-1},~~
\Delta(p/q)=(1-p^N/q^N)^{1/N},
$$
where $w(p/q,0)$ is yet arbitrary. The weight function $V$ satisfies the
three-dimensional star-star relation \cite{BB1,Hu1}
\begin{equation}
{\bar{V}(a|efg|bcd|h)\over V(a|efg|bcd|h)}={w(z,c-h-g+b)s(g+h,g-b)\over
 w(z,e-a-a+f)s(a+d,a-f)},
\end{equation}
where
$$
\bar{V}(a|efg|bcd|h) ~~~~~~~~~~~~~~~~~~~~~~~~~~~~~~~~~~~~~~~~~~~~~~~~~~~~~~~
{}~~~~~~~~~~~~
$$$$
=w_{q'q}(d-h-f+b)w^{-1}_{q'q}(e-c-a+g)s(c,g-a) ~~~
$$$$
{}~~~~ \times s(h,f-b)\sum^{N-1}_{\sigma=0}w^{-1}_{p'q}(\sigma-f+b)w_{pq}
(\sigma-a+g)
$$
\begin{equation} \label{2.3}
\times w_{q'p}(e-c-\sigma) \widetilde{w}_{p'q'}(\sigma-d+h)s(-\sigma,a-c-f+h),
\end{equation}
$$
z=e^{-i\pi/N}(\Gamma(p,p',q,q'))^{1/N},~\Gamma(p,p',q,q')=-{(p^N-q^N)(p'^N-q'^N)
\over(p^N-q'^N)(p'^N-q^N)}.
$$

\section{The derived of the weight function $R$ from Bazhanov-Baxter model}

Set
$$
x_1=q,~x_2=q',~x_3=p,~x_4=p',
$$
\begin{equation}
x^N_i-x^N_j=x^N_{ij},~i<j,~i,j=1,2,3,4.
\end{equation}
Following the notations in (\ref{2}), the weight function $V$ of the
Baxter-Bazhanov model can be written into the form
$$
\noindent
V(a|efg|bcd|h)                             \label{3.1}
{}~~~~~~~~~~~~~~~~~~~~~~~~~~~~~~~~~~~~~~~~~~~~~~~~~~~~~~~~~~~~~~~~~~~~~~~~~~~~~~~~
$$$$
={w(x_4,x_{34},x_3|e-c-d+h)s(g,a-g-f+b)\Phi(a-g)\over w(x_4,x_{34},x_3|a-g-f+b)
s(c-h,h-d)\Phi(b-f)} ~~~~~~~~~~~~~~~~
$$
\begin{equation}
{}~~~~~~~\times \Bigg\{ \sum^{N-1}_{\sigma=0}{w(_3,x_{13},x_1|\sigma+d-h)
w(x_4,x_{24},x_2|\sigma+a-g)s(\sigma,b+c-g-h)\over w(x_4,x_{14},x_1|\sigma+e-c)
w(x_3,x_{23},\omega x_2|\sigma+f-b)}\Bigg\}_0,
\end{equation}
where  the subscript "0" after the curly brackets indicates that the
expression
 in the braces is divided by itself with the zero exterior spins and  we have
 used the property
\begin{equation}                                \label{3.2}
w(x,y,z|l)w(z,\omega^{1/2}y,\omega x|-l)\Phi(l)=1,\quad l\in Z_N,
\end{equation}
and $\Phi(l)$ is given by (\ref{5}). Let
\begin{equation}
c=h,~~~d=e.
\end{equation}
Then the above weight function $V$ can be viewed as a Boltzmann weight function
 with the spins interacting round the triangular prism as in Fig. 2.

\begin{picture}(300.0,135.0)(-10,0)
\thicklines
\put(180,5){\circle*{15}}
\put(180,5){\line(1,0){80.0}}
\put(260,5){\circle*{15}}
\put(219.8,77.6){\circle*{15}}
\put(180,5){\line(1,2){35.8}}
\put(260,5){\line(-1,2){35.8}}
\put(180,5){\line(-2,1){71.5}}
\put(108.5,40.8){\circle*{15}}
\put(108.5,40.8){\line(1,2){35.8}}
\put(148.3,113.4){\circle*{15}}
\put(188.5,40.8){\circle*{15}}
\put(219.8,77.6){\line(-2,1){71.5}}
\multiput(108.5,40.8)(20,0){4}{\line(1,0){10.0}}
\multiput(260,5)(-17.9,8.9){4}{\line(-2,1){10.0}}
\multiput(188.5,40.8)(-8.9,17.9){4}{\line(-1,2){10.0}}
\put(164, 0){a}
\put(273,0){e}
\put(92.5,45.8){g}
\put(171.5,45.8){h}
\put(231.8,82.6){f}
\put(160,118.4){b}
\end{picture}

\bigskip
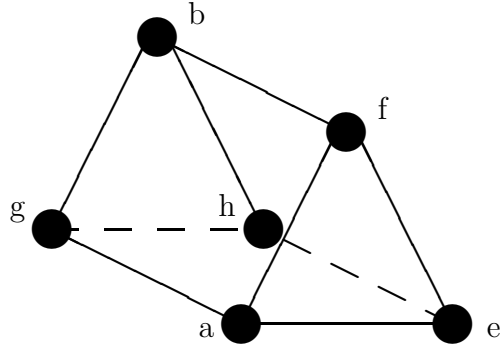
\centerline{Fig.2. {\it Arrangement of the spins $a,b,e,f,g,h$ on the
 elementary triangular prism}}
\bigskip
\noindent
Furthermore, when we make the choice of
\begin{equation}                \label{3.5}
 a+b=f+g,
\end{equation}
this figure can be denoted by Fig. 3.

\begin{picture}(300.0,100.0)(-20,+37)
\thicklines
\put(108.5,40.8){\circle*{15}}
\put(108.5,40.8){\line(1,0){80.0}}
\put(188.5,40.8){\circle*{15}}
\put(188.5,40.8){\line(-1,2){35.8}}
\put(148.3,113.4){\circle*{15}}
\put(108.5,40.8){\line(1,2){35.8}}
\put(219.8,77.6){\circle*{15}}
\put(219.8,77.6){\line(-2,1){71.5}}
\multiput(188.5,40.8)(-17.9,8.9){4}{\line(-2,1){10.0}}
\put(128.5,70.8){\line(-2,1){20.0}}
\put(108.5,80.8){\circle*{15}}
\put(92.5,35.8){g}
\put(198.5,35.8){h}
\put(158,118.4){b}
\put(94.5,85.8){e}
\put(229,67){f}
\end{picture}

\bigskip
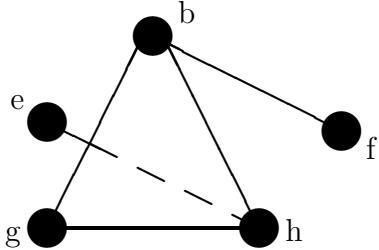
\centerline{Fig.3. \it {Arrangement of the spins $b,e,f,g,h$ corresponding to
  Eq. $($\ref{3.5}$)$}}
\bigskip
{}From the three-dimensional  star-star relation we know that the weight
function
 $V(a|efg|bcd|h)$ is the same with $\bar{V}(a|efg|bcd|h)$ modulo the factor
 $\omega^{(g-b)(b-f+h-e)}$ in this case. And the function $V$ can be written as
$$
V(a|efg|bcd|h)~~~~~~~~~~~~~~~~~~~~~~~~~~~~~~~~~~~~~~~~~~~~~~~~~~~~~~~~~~~~~~~~
{}~~~~~~~~~~~~~
$$
\begin{equation}                      \label{3.7}
{}~~~~~~=\Bigg\{ \sum^{N-1}_{\sigma=0}{w(x_3,x_{13},x_1|\sigma+e-h)
w(x_4,x_{24},x_2|\sigma+f-b)s(\sigma,b-g)\over w(x_4,x_{14},x_1|\sigma+e-h)
w(x_3,x_{23},\omega x_2|\sigma+f-b)}\Bigg\}_0,
\end{equation}
Now we make the transformation:
$$
b\longrightarrow -h,~e\longrightarrow c,~f\longrightarrow -f,
$$
\begin{equation}                                  \label{tra}
g\longrightarrow a-b-e,~~h\longrightarrow a,
\end{equation}
$$
V(a|efg|bcd|h) \longrightarrow w(a|efg|bcd|h),
$$
which means that the plane (bgh) in Fig.3 is substituted by the plane (aehb)
in Fig.4. Setting
\begin{equation}
x_4=0,~~x_1=x_{14},~~x_2=x_{24},
\end{equation}
we have that
$$
w(a|efg|bcd|h)~~~~~~~~~~~~~~~~~~~~~~~~~~~~~~~~~~~~~~~~~~~~~~~~~~~~~~~~~~~~~~~
{}~~~~~~~
$$
\begin{equation} \label{ww}
{}~~~~=\omega^{(h-f)(a-b-e+h)}\Bigg\{\sum^{N-1}_{\sigma=0}
{w(x_3,x_{13},x_1|-\sigma-a+c+f-h)\over w(x_3,x_{23},\omega x_2|-\sigma)
s(\sigma,-a+b+e-h)}\Bigg\}_0.
\end{equation}

\begin{picture}(300.0,150.0)(-10,0)

\put(128.5,30.8){\line(1,1){55.0}}
\multiput(208.5,30.8)(15,15){4}{\line(1,1){10.0}}

\thicklines
\put(180,5){\circle*{15}}
\put(180,5){\line(1,0){80.0}}
\put(260,5){\circle*{15}}
\put(180,5){\line(0,1){80.0}}
\put(180,85){\circle*{15}}
\put(180,85){\line(1,0){80.0}}
\put(260,85){\circle*{15}}
\put(260,5){\line(0,1){80.0}}
\put(180,5){\line(-2,1){54}}
\put(128.5,30.8){\circle*{15}}
\put(128.5,30.8){\line(0,1){80.0}}
\put(128.5,110.8){\circle*{15}}
\put(180,85){\line(-2,1){54}}
\put(128.5,110.8){\line(1,0){80.0}}
\put(208.5,110.8){\circle*{15}}
\put(260,85){\line(-2,1){54}}
\put(208.5,30.8){\circle*{15}}
\multiput(260,5)(-17.9,8.9){3}{\line(-2,1){10.0}}
\multiput(128.5,30.8)(20,0){4}{\line(1,0){10.0}}
\multiput(208.5,30.8)(0,20){4}{\line(0,1){10.0}}
\put(162, 0){e}
\put(162,80){a}
\put(275,0){d}
\put(275,80){f}
\put(112.5,35.8){c}
\put(111.5,115.8){g}
\put(223.5,115.8){b}
\put(223.5,31.8){h}
\put(180,85){\line(6,5){35}}
\multiput(180,5)(18,15){2}{\line(6,5){10.0}}

\put(180,8){::}
\put(180,16){::::}
\put(180,25){:::::::::}
\put(180,33){:::::::::}
\put(180,42){:::::::::}
\put(180,50){:::::::::}
\put(180,58){:::::::::}
\put(180,62){:::::::::}
\put(180,70){:::::::::}
\put(180,78){:::::::::}
\put(180,83){::::::;::}
\put(180,91){~~~:::::}
\put(180,99){~~~~~~::}

\end{picture}

\bigskip
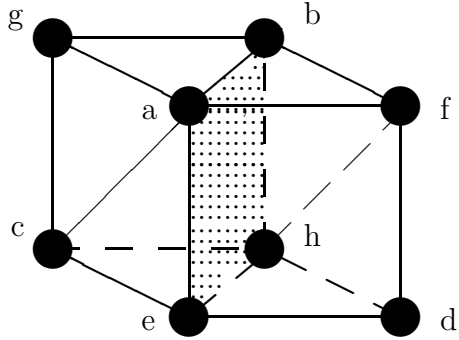
\centerline{Fig.4. {\it Arrangement of the spins after transforming
 $($\ref{tra}$)$}}
\bigskip
\noindent
The restricted star-triangle relation of the Bazhanov-Baxter model has the
 form \cite{BB2,Hu2,PHD,Chin,Hu3}
\begin{equation}          \label{str}
\sum^N_{l=1}{w_{pR(r)}(n-l)\over w_{qr}(l-m)w_{pq}(k-l)}=\rho (pqr){w_{pR(q)}
(n-m)w_{R^{-1}(q)r}(k-n)\over w_{pr}(k-m)}
\end{equation}
where
$\rho(pqr)$ is a scalar factor and
\begin{equation}
R(a_p,b_p,c_p,d_p)=(b_p,\omega a_p,d_p,c_p),~w_{pq}(n)
=w(\omega^{-1}c_pb_q,d_pa_q,b_pc_q|n)
\end{equation}
with $a_p=d_r=0$. The vectors $(a_p,b_p,c_p,d_p) ~etc$ satisfy
\begin{equation}
a^N_p+k'b^N_p=kd^N_p,~k'a^N_p+b^N_p=kc^N_p
\end{equation}
with $k^2+k'^2=1$. By using the above restricted star-triangle relation,
 Eq. (\ref{ww}) can be transformed into the form:
$$
w(a|efg|bcd|h)=\omega^{(h-f)(a-b-e+h)} \times~~~~~~~~~~~~~~~~~~~~~~~
{}~~~~~~~~~~~~~~~~~~~~~~~~~~
$$
\begin{equation}          \label{29}
{}~~~~~~~~~~~~~~~~~~~~\times{w(p_2,p_{12},p_1|a-b-e+h)w(p_4,p_{34},p_3|-a+c+f-h)
\over w(p_6,p_{56},p_5|-b+c-e+f)}
\end{equation}
where
$$
p_1=\omega x_{13}x_2,~p_2=x_1x_{23},~p_3=\omega x_1x_3,
$$
\begin{equation}
p_4=\omega x_2x_{3},~p_5=\omega x_{13}x_3,~p_6=x_{23}x_3
\end{equation}
with $p^N_{ij}=p^N_i-p^N_j$ for $i<j$. This is just the relation (2.9) of the
Ref. \cite{Man}. It can be written as the form of the equation (\ref{1}) by
making a proper choice of  the spin variables. So we get the connection of
 the Bazhanov-Baxter model and the one proposed by Mangazeev $et~al$.

\section{Conclusion and Remarks}

As the above discussion, we have obtained the $N$-state spin integrable model
 on a three-dimensional lattice with the spins interacting round each
 elementary cube of the lattice, proposed by Mangazeev, Sergeev and Stroganov,
 from the Bazhanov-Baxter model. The key point in the above derivation is
 that the spectrums  in  Eq. (1) have the relation  $p_6/p_5=p_2p_4/(p_1p_3)$
 and the spin variable appeared in the denominator of the right hand side
term of Eq. (1) is the sum of that appeared in the numerator, which is similar
 to the property of the restricted star-triangle relation of the
Bazhanov-Baxter model. We know that the tetrahedron equation plays an
 important role in the IRC model. And it is proved that the weight function
 $V(a|efg|bcd|h)$ of the Eq. (\ref{2.1}) satisfies the tetrahedron equation
 by using the method of the spherical trigonometry parametrization where
some additional multipliers  was introduced \cite{Kae1}. But the weight
function (1)  cannot be derived from the result of this parametrization when
 $N>2$ \cite{Man}. It would be an interesting question to find a new
 parametrization for weight function $V(a|efg|bcd|h)$ of the Bazhanov-Baxter
 model which satisfies the tetrahedron equation and contains the relation
 (\ref{29}) as a limited case.

\section*{Acknowledgment}
The author would like to thank Prof. K. Wu for his encouragements and
 Dr. R. Q. Lau, Q. P. Liu for  the interesting discussions. I am also
 grateful Dr. H. Yan for taking me many papers and preprints from Kyoto.

\end{document}